\documentclass[conference]{IEEEtran}

\IEEEoverridecommandlockouts

\usepackage{fancyhdr}

\usepackage{eso-pic}

\ifCLASSINFOpdf
\else
\fi

\usepackage{algpseudocode}
\usepackage[section]{algorithm}

\usepackage{amsmath}
\usepackage{amssymb}
\usepackage{mathtools}
\usepackage{graphicx}
\usepackage{subfigure}
\usepackage{caption}
\usepackage{cases}
\newcommand\bovermat[2]{%
  \makebox[0pt][l]{$\smash{\overbrace{\phantom{%
    \begin{matrix}#2\end{matrix}}}^{\text{#1}}}$}#2}

\begin{document}
\title{Distributed Power Control in Multiuser MIMO Networks with Optimal Linear Precoding}\author{\IEEEauthorblockN{Peyman Siyari}
\IEEEauthorblockA{Department
of Electrical Engineering\\Amirkabir University of Technology\\Tehran,
Iran\\Email: psiyari@aut.ac.ir}
\and
\IEEEauthorblockN{Hassan Aghaeinia}
\IEEEauthorblockA{Department
of Electrical Engineering\\Amirkabir University of Technology\\Tehran,
Iran\\Email: aghaeini@aut.ac.ir}%
}
\maketitle
\begin{abstract}
Contractive interference functions introduced by  Feyzmahdavian et al.\@ is the newest approach in the analysis and design of distributed power control laws. This approach can be extended to several cases of distributed power control. One of the distributed power control scenarios wherein the contractive interference functions have not been employed is the power control in MIMO systems. In this paper, this scenario will be analyzed. In addition, the optimal linear precoder is employed in each user to achieve maximum point-to-point information rate. In our approach, we use the same amount of signaling as the previous methods did. However, we show that the uniqueness of Nash equilibria is more probable in our approach, suggesting that our proposed method improves the convergence performance of distributed power control in MIMO systems. We also show that the proposed power control algorithm can be implemented asynchronously, which gives a noticeable flexibility to our algorithm given the practical communication limitations.\end{abstract}

\IEEEpeerreviewmaketitle

\thispagestyle{fancy}
\section{Introduction}
\IEEEPARstart{I}{n} many wireless networks, such as ad-hoc networks or cognitive radio systems, there are multiple independent users that share a common communication environment. The interference channel is the most relevant mathematical model used to analyze the interaction of users in such networks. In order to calculate an inner bound for the capacity of the interference channel, the simplifying and pragmatic assumptions such as neglecting the use of interference cancellation techniques are common in the analysis of interference channels \cite{scutari1}. As these assumptions suggest, the analysis of the capacity of the interference channel simplifies to the analysis of power consumption of the users in network. Motivated by this fact, there are several studies about the power allocation of the users in an interference channel, which promise to reach the optimum capacity of interference channel under different scenarios (see \cite{yu1, yu2, yu3} and references therein). However, these solutions are based on either centralized approaches or heavy control signaling between the nodes which may not be of our interest in a network with the scarcity of bandwidth and power to the users. Therefore, distributed approaches become of particular interest, since they do not need that much coordination and signaling between the nodes, or at least the signaling is done only between each transmit-receive pair. 

One of the first analyses of a distributed power control law was done in \cite{yates}, where Yates et al. introduced the concept of standard functions that permits a general proof of the synchronous and totally asynchronous convergence of the iterative power control to a unique fixed point minimizing the total transmit power for each user. After the work of Yates, Feyzmahdavian et al.\@ in \cite{feyz} introduced a slight and clever variation of the standard functions, namely as contractive functions to introduce a more unified explanation of distributed power control frameworks. With contractive functions, the uncertainties about the standard functions such as the existence of fixed points and rates of convergence could be explained. Moreover, the concept of contractive interference functions was extended by the original authors themselves to several cases of distributed power control.

The \lq\lq performance coupling\rq\rq among the users is the main challenge of using distributed approaches because the increase of one user's performance comes at the price of performance degradation of others.
One of the most applicable tools in distributed optimization is the use of game theoretical concepts. In fact, we are interested in modeling the distributed power control in an interference channel as a non-cooperative game, where every user is a player that competes against the others by choosing the power allocation that maximizes its own information rate. A Nash equilibrium is achieved when no player can unilaterally increase its own information rate given the current strategies of the others \cite{funden}. Analysis of power control in interference channel using non-cooperative game theory was first done in \cite{yu3}, where Yu, et al.\@ solved the power control problem in Digital Subscriber Line (DSL) systems by modeling the system as a gaussian frequency-selective interference channel and performing Iterative Water-Filling Algorithm (IWFA). Since then, a number of studies have been done to extend this original idea, and finally the studies of Scutari et al.\@ brought a unified view of distributed IWFA in many scenarios such as multiuser MIMO and cognitive radio systems (see \cite{sc1, sc3}).

In this paper, distributed power control in multiuser MIMO systems, is investigated with the help of game theory and contractive interference functions. 
Our approach is compared with the state-of-the-art method used in \cite{sc3}, where the authors give a different interpretation of water-filling by using it as a contraction mapping. We show that with the use of the same amount of signaling as in the previous papers, the uniqueness of Nash equilibria is more probable in our approach. Furthermore, we show that our distributed power control algorithm is flexible enough to deal with various practical limitations of communications networks such as communication delays.

The rest of this paper is organized as follows. In section II, the general system model is described. In section III, the power control problem is formulated as a non-cooperative game, and the elements of the game are determined. In section IV, a brief introduction of contraction mappings and contractive functions is presented. After that, the analysis of uniqueness of Nash equilibrium is done, and the condition that guarantees the uniqueness of NE in power control game is established. Next, the design of distributed power control algorithms is mentioned in section V. In section VI, the computer simulations verify our theoretical analysis in previous sections. Finally, section VII concludes the paper.
\section{System Model}
In this paper, we employ a multiuser MIMO channel as our system model. There are Q transmit-receive pairs, and each pair is a user. The $q$th transmitter has $N_{T_q}$ transmit antennas and the $q$th receiver has $N_{R_q}$ receive antennas. The received signal at each receiver interfered by the other $Q-1$ transmitters. As we do not allow any multiuser encoding or coordination among the users, the interference at each receiver is treated as an additive component. Hence, the received signal at the $q$th receiver $y_q$ is
\begin{align}
\label{sysmodel}
y_q = H_{qq}x_q+\sum_{r \neq q}H_{rq}x_r+n_q,
~~\forall q\in\{1,...,Q\}
\end{align}
where $x_q \in C^{N_{T_q}}$ is the complex transmitted signal from the $q$th transmitter, $y_q \in C^{N_{R_q}}$, $H_{rq} \in C^{N_{R_q}\times N_{T_r}}$ is the complex channel matrix between the $r$th transmitter and $q$th receiver, and $n_q$ is the additive white Gaussian noise (AWGN) vector with the covariance matrix $R_{n_q} = N_{0_q}I_{N_{R_q}\times N_{R_q}}$ where the scalar value $N_{0_q}$ is the power of the noise and $I$ is the identity matrix. Hence, the vector $\sum_{r \neq q}H_{rq}x_r$ is the multiuser interference (MUI) at the $q$th receiver. Given $\mathcal{E}\left\{ x_qx_q^H\right\}$ as the covariance matrix of the vector $x_q$, the power constraint for each transmitter is
\begin{equation}
\label{powerconstraint}
\mathcal{E}\left\{ ||x_q||_2\right\} = Tr\left(\mathcal{E}\left\{ x_qx_q^H\right\}\right)\le P_q,
\end{equation}
where $||.||_2$ is the second vector norm, $Tr(.)$ is the trace operator, and $P_q$ is a scalar value that represents the total amount of power that a transmitter can distribute between its antennas.  Assuming that the vector $p_q = [p_q^1, ...,p_q^{N_{T_q}} ]^T = diag\left(\mathcal{E}\left\{ x_qx_q^H\right\}\right)$ indicates the power of the transmitted signal from the $q$th user, the power constraint for the $q$th user can alternatively be shown as:
\begin{equation}
\sum_{i = 1}^{N_{T_q}}p_q^i \le P_q.
\end{equation}
\begin{figure}
\centerline{
\includegraphics[scale = 0.3]{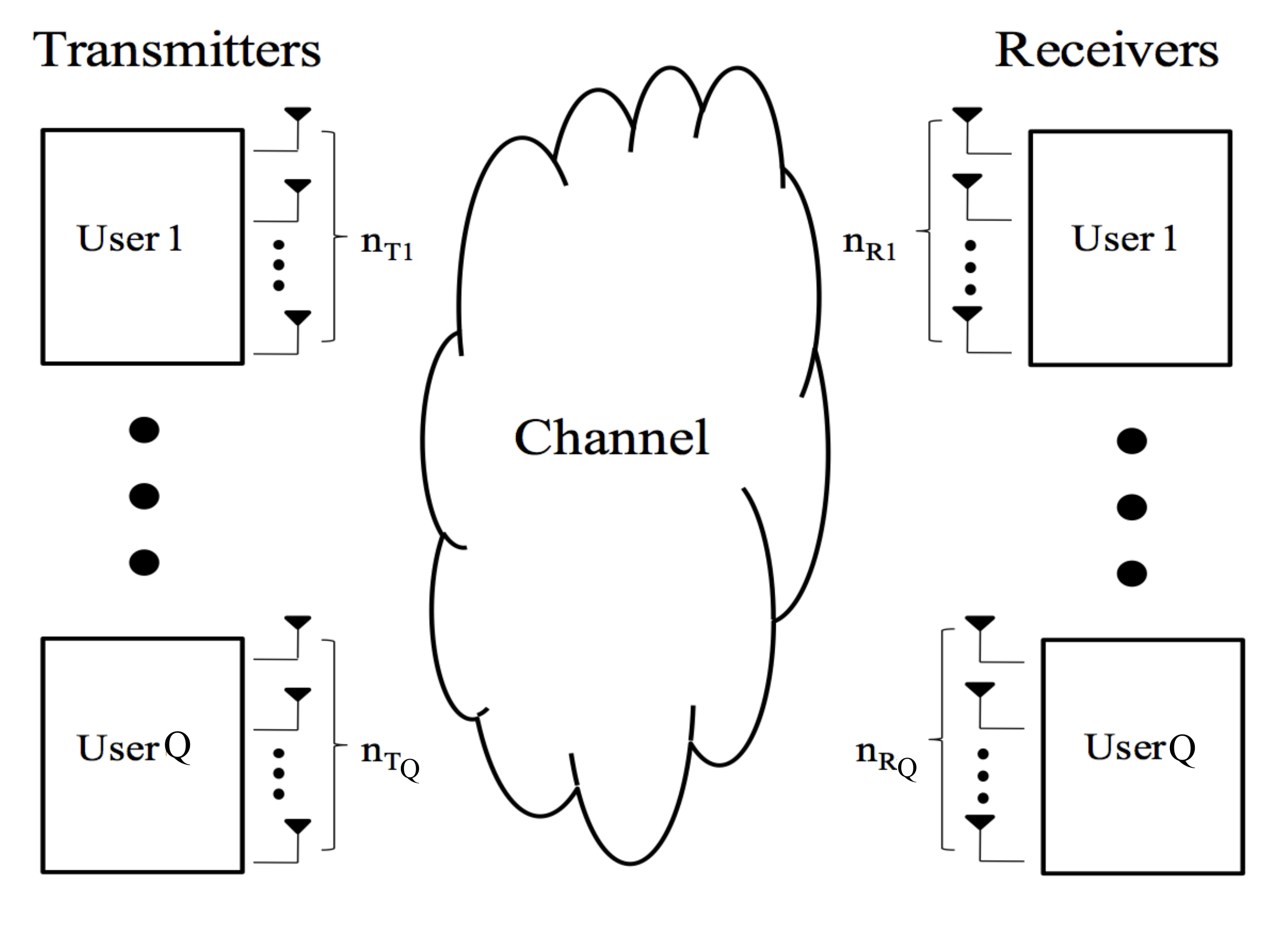}
}
\caption{\small System Model}
\end{figure}
\section{Problem Formulation}
\subsubsection*{Using Channel State Information (CSI)}
We assume that each transmitter knows the channel between itself and its receiver pair \footnote{This assumption imposes no overhead more than the previous related work (compare with \cite{sc3}).}. It can be shown that the knowledge of CSI (in i.i.d channels) at the transmitter can significantly improve the capacity of channel \cite{paulraj}. The knowledge of CSI can be exploited by designing an appropriate precoder. With the knowledge of CSI, the capacity-achieving scheme is to use singular-value decomposition to diagonalize the channel. The singular-value decomposition of $H_{qq}$ yields
\begin{equation}
H_{qq} = U_q \Sigma_q V_q^H,
\end{equation}
where $U_q$ and $V_q$ are unitary matrices and $\Sigma_q$ is the diagonal matrix of singular values of $H_{qq}$. We assume the vector $diag(\Sigma_{q}) = \sigma_q = [\sigma_q^1, ..., \sigma_q^{\nu_{q}}]^T$ with $\nu_q = min(N_{T_q},N_{R_q})$ is the vector of the singular values of $H_{qq}$. Using $V_q$ as the precoder matrix and $U_q^H$ as the decoding matrix, the received signal at the $q$th receiver would be
\begin{equation}
\label{svd}
\tilde{y}_q = \Sigma_{q}x_q+\sum_{r \neq q}U_q^HH_{rq}V_r x_r+U_q^Hn_q,
\end{equation}
where $\tilde{y}_q = U_q^Hy_q$. Since $U$ and $V$ are invertible, knowing $\tilde{y}_q$ is the same as knowing $y_q$. Furthermore, the noise vector's statistics do not change when multiplied by $U_q^H$, since $U_q$ is a unitary matrix and $n_q$ has independent Gaussian elements \cite[Chapter 3]{duman}. We take $\tilde{H}_{rq} = U_q^HH_{rq}V_r$ as an $N_{R_q} \times N_{T_r}$ matrix. The element in the $i$th row and $j$th column of $\tilde{H}_{rq}$ is shown by the operator $\left[\tilde{H}_{rq}\right]_{i,j}$. Assuming no correlation between the transmit and receive antennas, the capacity of the MIMO channel between the $q$th transmit-receive pair would be:
\begin{equation}
R_q = \sum_{i=1}^{\nu_q} \log\left(1+\frac{p^i_q}{c^i_q}\right),
\end{equation}
where $c^i_q$ --the normalized interference plus noise of the $q$th user-- is written as
\begin{equation}
c^i_q = \sum_{r \neq q}\left\{\sum_{j = 1}^{{N_T}_r}\frac{\left|\left[\tilde{H}_{rq}\right]_{i,j}\right|^2p^j_r}{|\sigma_q^i|^2}\right\}+\frac{N_{0_q}}{|\sigma_q^i|^2},~~i\in\{1,...,\nu_q \}.
\end{equation}
\subsubsection*{Game Formulation}
As each user myopically chooses the best strategy for itself, we formulate this scenario as a non-cooperative game, in which the best strategy of each user is
\begin{align}
\max_{p_q}~R_q
\\
\text{s.t.}~\sum_{i = 1}^{N_{T_q}}p^i_q \le P_q
\notag.
\end{align}
Therefore, in this game, the utility function of each player (or user) is its capacity and the strategy of each player is choosing the best power allocation. Since the capacity of each user is a concave function, the maximum capacity is achieved by performing water-filling which can be shown as
\begin{equation}
\label{water}
{p^i_q}^* = \left(\mu_q-c^i_q\right)^+,
\end{equation}
where $(...)^+ = max(...~,~0)$ and $\mu_q$ is a positive value that satisfies the power constraint.

The existence of Nash equilibrium in this game can be proven by showing that the strategy set of each user is non-empty, compact and convex subset of some Euclidean real space $\mathcal R^{N_{T_q}}$, and the utility function of each user is a continuous and quasi-concave function of its own power vector. The detailed proof for the existence of NE is straight-forward and skipped due to space limitation. The uniqueness of equilibrium is investigated in the next section.
\section{Uniqueness of Nash Equilibrium}
So far, we have formulated a non-cooperative power control game in a multi-user MIMO system, in which every transmitter performs water-filling in allocating the power between its antennas in order to achieve the maximum capacity. In this section, we introduce a new characterization of MIMO power control. First, we review fundamentals of fixed-point theory as the basis of our analyses.
\subsubsection*{Fixed-Point Theory}
Consider the following iteration:
\begin{equation}
\label{fix}
x(k+1) = T\left(x(k)\right),~k=1,2,...,
\end{equation}
where $T$ is a mapping from a subset $X$ of $\mathcal R^K$ to itself and $k$ indicates the index of iterations. If T is continuous and
\begin{equation}
||T(x)-T(y)|| \le c||x-y||~,~\forall x,y \in X
\end{equation}
where $||.||$ is some norm in $X$ and $c \in [0,1)$, then the mapping $T$ is a contraction mapping with $c$ as the contraction modulus, and sequence $\left\{x(n)\right\}$ generated by the iterations in \eqref{fix} converges to the fixed point $x^*$ \cite[Chapter 13]{feyz13}.

We consider the mapping $T$ as the water-filling operator written in equation \eqref{water}. In the following, the contractivity of the mapping $T$ is analyzed. 
\subsubsection*{Contractive Functions}
A function $I: \mathcal R^K_+ \rightarrow \mathcal R^K_+$ (with $\mathcal{R}^K_+ = \{p|p>0, p\in \mathcal{R}\}$) is said to be contractive if it, for all $p \ge 0$ satisfies \cite{feyz}:
\begin{itemize}
\item{}
Positivity: $I(p)\ge 0$
\item{}
Monotonicity: if $p \ge p'$, then $I(p) \ge I(p')$
\item{}
Contractivity: There exists a constant $c \in [0,1)$ --namely as the contractivity modulus-- and a vector $v > 0$ such that $I(p+\epsilon v) \le I(p) + c\epsilon v,~~\forall \epsilon > 0$.
\end{itemize}
The following results have been deriven about contractive functions \cite{feyz}:
\newtheorem{proposition}{\bf{Proposition}}
\begin{proposition}
If a function $I: \mathcal R^K_+ \rightarrow \mathcal R^K_+$ is contractive with $c\in[0,1)$ as the contractivity modulus, then
\begin{enumerate}
\item{}
It is a contraction mapping with maximum norm and $c\in [0,1)$ as the contraction modulus, then it has a unique fixed point $p^*$, and the same iteration for $I(p)$ according to \eqref{fix} will converge to $p^*$.
\item{}
The following function is also contractive with the same modulus $c \in [0,1)$:
\begin{equation}
I^q(p) = max\left\{p_{min}, min\{p_{max}, I(p)\}\right\},
\end{equation}
where $p_{min}$ and $p_{max}$ may be the minumum and miaximum constraints on the power allocation, respectively.
\end{enumerate}
\hfill $\square$
\end{proposition}
With these results, we are now ready to present a new theorem on the convergence of distributed power control in our scenario.
\newtheorem{theorem}{\bf{Theorem}}
\begin{theorem}
The MIMO power control game defined in the previous section with the players' strategies written as \eqref{water} has a unique Nash equilibrium if $\forall (i,q) \in \{1,..., \max_q\{\nu_{q}\}\} \times\{1,..,Q\}$ we have
\begin{equation}
\label{th1}
\sum_{j=1}^{N_T}\max_{(i,q)}\left\{\sum_{r\neq q}    \frac{\left|\left[\tilde{H}_{rq}\right]_{i,j}\right|^2}{|\sigma_q^i|^2}\right\}<1,
\end{equation}
or alternatively
\begin{equation}
\label{th2}
\sum_{j=1}^{N_T}\max_{(i,q)}\left\{\sum_{r\neq q}    \frac{\left|\left[\tilde{H}_{qr}\right]_{i,j}\right|^2}{|\sigma_q^i|^2}\right\}<1.
\end{equation}
\begin{IEEEproof}
As we are analyzing a power control game, we should analyze the strategies of the users in order to find the Nash equilibria of the game.
Assuming $p = [p_1, p_2, ..., p_q, ..., p_Q]^T$ with $p_q = [p_q^1, ...,p_q^{N_{T_q}} ]^T = diag\left(\mathcal{E}\left\{ x_qx_q^H\right\}\right)$, we first prove that the interference function $I(p) = [I_1(p), ..., I_Q(p)]^T$ is contractive, where $I_q(p) = [I^1_q(p),...,I^{\nu_q}_q(p)]^T$ and $I^i_q(p) = \mu_q - c^i_q$. Next, we use the second item in proposition 1 to prove that the water-filling operator is also a contractive function.

\begin{figure*}[!t]
\small
\begin{equation}
\label{intmat}
\hspace{-7mm}
M = 
\begin{matrix}
 \begin{bmatrix}
\bovermat{$N_{T_1}$}{&~~~~~~~~~&~~~~~~~~~~~~~~~~~&}
\\[-2.3em]
0 & \cdots & 0 &
\left|\frac{\left[\tilde{H}_{21}\right]_{1,1}}{\sigma^1_1}\right|^2 & \cdots & \left|\frac{\left[\tilde{H}_{21}\right]_{1,N_{T_2}}}{\sigma^1_1}\right|^2 & \left|\frac{\left[\tilde{H}_{31}\right]_{1,1}}{\sigma^1_1}\right|^2 & \cdots & \left|\frac{\left[\tilde{H}_{Q1}\right]_{1,N_{T_Q}}}{\sigma^1_1}\right|^2 
\\[1.5em]
0 & ~~~~\cdots~~~~ & 0 & \left|\frac{\left[\tilde{H}_{21}\right]_{2,1}}{\sigma^2_1}\right|^2 & \cdots & \left|\frac{\left[\tilde{H}_{21}\right]_{2,N_{T_2}}}{\sigma^2_1}\right|^2 & \left|\frac{\left[\tilde{H}_{31}\right]_{2,1}}{\sigma^2_1}\right|^2  & \cdots & \left|\frac{\left[\tilde{H}_{Q1}\right]_{2,N_{T_Q}}}{\sigma^2_1}\right|^2 
\\[0.8em]
\vdots&&\vdots&\vdots&&\vdots&\vdots&&\vdots
\\[0.8em]
0 & ~~~~\cdots~~~~ & 0& \left|\frac{\left[\tilde{H}_{21}\right]_{\nu_1,1}}{\sigma^{\nu_1}_1}\right|^2 & \cdots & \left|\frac{\left[\tilde{H}_{21}\right]_{\nu_1,N_{T_2}}}{\sigma^{\nu_1}_1}\right|^2 & \left|\frac{\left[\tilde{H}_{31}\right]_{\nu_1,1}}{\sigma^{\nu_1}_1}\right|^2  & \cdots & \left|\frac{\left[\tilde{H}_{Q1}\right]_{\nu_1,N_{T_Q}}}{\sigma^{\nu_1}_1}\right|^2 
\\[3em]
&&&\bovermat{$N_{T_2}$}{&~~~~~~~&~~~~~~~~~~~~~~~~~}
\\[-3.5em]
\\
\left|\frac{\left[\tilde{H}_{12}\right]_{1,1}}{\sigma^{1}_2}\right|^2 & \cdots & \left|\frac{\left[\tilde{H}_{12}\right]_{1,N_{T_1}}}{\sigma^{1}_2}\right|^2
& 0 & \cdots & 0 & \cdots & \cdots & \left|\frac{\left[\tilde{H}_{Q2}\right]_{1,N_{T_Q}}}{\sigma^{1}_2}\right|^2
\\[0.8em]
\vdots&&\vdots&\vdots&&\vdots&\vdots&&\vdots
\\[0.8em]
\left|\frac{\left[\tilde{H}_{12}\right]_{\nu_2,1}}{\sigma^{\nu_2}_2}\right|^2 & \cdots & \left|\frac{\left[\tilde{H}_{12}\right]_{\nu_2,N_{T_1}}}{\sigma^{\nu_2}_2}\right|^2
& 0 & \cdots & 0 & \cdots & \cdots & \left|\frac{\left[\tilde{H}_{Q2}\right]_{\nu_2,N_{T_Q}}}{\sigma^{\nu_2}_2}\right|^2
\\[0.8em]
\vdots&&\vdots&\vdots&&\vdots&\vdots&&\vdots
\\[0.8em]
&&&&&&\bovermat{$N_{T_Q}$}{&~~~&~~~~~~~~~~~~~~~~~}
\\[-3.5em]
\\
\left|\frac{\left[\tilde{H}_{1Q}\right]_{1,1}}{\sigma^{1}_Q}\right|^2 & \cdots & \left|\frac{\left[\tilde{H}_{1Q}\right]_{1,N_{T_1}}}{\sigma^{1}_Q}\right|^2
&
\left|\frac{\left[\tilde{H}_{2Q}\right]_{1,1}}{\sigma^{1}_Q}\right|^2 &  \cdots & \cdots
& 0 & \cdots & 0
\\[0.8em]
\vdots&&\vdots&\vdots&&\vdots&\vdots&&\vdots
\\[0.8em]
\left|\frac{\left[\tilde{H}_{1Q}\right]_{\nu_Q,1}}{\sigma^{\nu_Q}_Q}\right|^2 & \cdots & \left|\frac{\left[\tilde{H}_{1Q}\right]_{\nu_Q,N_{T_1}}}{\sigma^{\nu_Q}_Q}\right|^2
&
\left|\frac{\left[\tilde{H}_{2Q}\right]_{\nu_Q,1}}{\sigma^{\nu_Q}_Q}\right|^2 & \cdots  & \cdots
& 0 & \cdots & 0
  \end{bmatrix}
 \begin{aligned}
\\[-3em]
\\
\left.
\begin{matrix}
\\ \\ \\ \\ \\ \\ \\ \\ \\ \\
  \end{matrix} \right\} \nu_1
 \\ \\
\left.
\begin{matrix}
\\ \\ \\ \\ \\ \\ \\
  \end{matrix} \right\}\nu_2
\\
\\[-1em]
\vdots~~~~
\\
\\[-.6em]
\left.
\begin{matrix}
\\ \\ \\ \\ \\ \\
  \end{matrix} \right\} \nu_Q
 \end{aligned}
\\
\\
\end{matrix}
\end{equation}
\hrulefill
\end{figure*}

The contractivity of the interference functions can be shown if we are able to show the vector $I(p)$ with a close-form representation. As we used the SVD precoder, there are $\nu_q = min (N_{T_q}, N_{R_q})$ parallel channels between each transmit-receive pair.
Assuming $N = [\mu_1-\frac{N_{0_1}}{|\sigma^1_1|}, ..., \mu_1-\frac{N_{0_1}}{|\sigma^{\nu_1}_1|},... ,\mu_Q-\frac{N_{0_Q}}{|\sigma^{\nu_Q}_Q|}]^T$, the interference function $I(p)$ is
\begin{equation}
I(p) = M.p+N,
\end{equation}
where $M$ is a $\{(\sum_{q=1}^Q\nu_q)\} \times \{(\sum_{q=1}^QN_{T_q})\}$ matrix which is shown in \eqref{intmat}.
Depending on the value of $\nu_q$ for all $q$, we present different proofs for contractivity.

{\bf 1) $\bf \nu_q = N_{T_q} = N_{R_q}$:}
In this case, the matrix $M$ is a square matrix. While checking the contractivity properties of $I(p)$, we conclude that the first two properties (i.e. positivity and monotonicity) are evident. For the third property (i.e. contractivity property), assuming a positive vector $v$, we have
\begin{equation}
\label{cont}
I(p+\epsilon v) = M.p +N + M.\epsilon v \le I(p) + ||M||_\infty^v \epsilon v,
\end{equation}
where $||.||_\infty^v$ is the weighted maximum norm given the vector $v > 0$. Therefore, if $||M||_\infty^v < 1$, the function $I(p)$ is a contractive function, and according to the properties of contractive functions, the iterative water-filling between the users will converge to a unique fixed point that is the Nash equilibrium of the power control game. It is known that for the non-negative square matrices (e.g. $M$), there exists a positive vector $v$ such that $||M||_\infty^v < 1$ if and only if $\rho(M) <1$ \cite{feyz13}. Both of these measures are difficult to evaluate in the practical cases because the designer has to have access to the matrix $M$ to predict the uniqueness of NE. The easiest verifiable conditions can be derived by choosing $v = 1$, which yields the inequality in \eqref{th1}, or equivalently $||M||_\infty < 1$. Since $\rho(M) = \rho(M^T)$, all of the aforementioned derivations about  a non-negative square matrix can be done for its transposed version (i.e. $M^T$). Doing so for $M^T$ yields the inequality in \eqref{th2}. The physical interpretation of \eqref{th1} and \eqref{th2} suggests that the sum of normalized interference produced by each user has to be less than one, or alternatively the sum of received normalized interference at each receiver should be less than one. This intuitive interpretation can be easily used by the designer depending on the availability of the value of normalized interference for each user.

{\bf 2) $\bf \nu_q = N_{T_q} < N_{R_q}$:} As \eqref{intmat} suggests, in this case, the matrix $M$ is still a square matrix and all of the aforementioned proof about the previous case is also true about this case.

{\bf 3) $\bf \nu_q = N_{T_q} > N_{R_q}$:} In this case, the matrix $M$ is not square and the proofs of the previous cases cannot be employed for this case. As we want to have a set of unified conditions for all the cases, a simple manipulation of the matrix $M$ can make it a square matrix for this case. This manipulation is done by adding several rows of zeroes to the matrix $M$ for the user --say $q$th user-- that has $N_{T_q} > N_{R_q}$. Adding zero rows is done until sum of the $q$th user's transmit antennas equals to the sum of both its receive antennas and the zero rows added for the $q$th user. We should note that by adding rows of zeroes for each user whose transmit antennas are more than its receive antennas, the whole measurements for the interference will not be affected. In fact, we assume that adding zero rows to $M$ is equal to adding a receive antenna that does not receive neither signal nor interference (i.e. it is turned off). Hence, the matrix $M$ is eventually square again and we can use the same proofs we used previously.
%
\end{IEEEproof}
\end{theorem}
\section{Algorithm Design}
In the previous section, it was proved that the distributed iterative waterfilling that each user performs is a contraction mapping with maximum norm. The immediate result of this proof is that the iterative water-filling algorithm can be done totally asynchronously in the sense of \cite{feyz13}. Let $T_q,~\forall q \in \{1,2,...,q\}$ be the set of times (or iteration numbers) when $q$th user updates its power allocation. While updating the power allocation,  maybe an outdated power radiated by the interfering links is used by a user in the calculation total interference, then the power allocation may not be according to the recent changes in the network. This usually happens due to the communication delay that may occur in the network. To model that, we assume that $\tau^q(n) = \{\tau_1^q(n), ...,\tau_Q^q(n)\}$ as the set of most recent times that the information from each user is received by the $q$th user at time $n$. Therefore, at each time $n \in T_q$ the $q$th transmitter performs waterfilling based on the $\tau^q(n)$ that is available at time $n$. With these definitions, we are ready to present our asynchronous power control algorithm for the iteration number $n$ which is as follows:

\begin{algorithm}[H]
\caption{Asynchronous Iterative Water-filling}
\label{alg}
Set~$p_q(0)$ such that 
$\sum_{i = 1}^{N_{T_q}}p^i_q \le P_q$
\begin{algorithmic}[1]
\For{n=1 to $it_{max}$}
\Repeat
~~~$\forall (i,q)\in\{1,...,\nu_q\}\times\{1,...,Q\}$
\State
$
{p^i_q(n+1)} =\left\{
\begin{array}{c l}     
\left(\mu_q(n)-c^i_q(n)\right)^+ & \text{if~}n\in T_q\\
    {p^i_q(n)} & \text{otherwise}
\end{array}\right.$,
\Until{Convergence}
\EndFor
\end{algorithmic}
\end{algorithm}
In the above algorithm, $it_{max}$ denotes the maximum number of iterations, and the fourth step can be replaced with a particular termination criterion. Furthermore, in the term $c^i_q(n)$, the most recent power updates of other users are available (according to $\tau_q(n)$). The special cases of asynchronous implementation include synchronous implementation (Jacobi algorithm \cite{feyz13}) and sequential implementation (Gauss-Seidel algorithm \cite{feyz13}), that are used in the previous state-of-the-art algorithms \cite{yu1, yu2, yu3}. For the Jacobi Algorithm we have \begin{align}
T_q = \{1, 2, ..., it_{max}\}
,\ 
\tau^q(n) = \{n,...,n\}
\end{align}
which means at each iteration all of the users simultaneously update their power allocation, and for Gauss-Seidel we have 
\begin{align}
T_q = \{q, q+Q, q+2Q, ..., q+\left(\frac{it_{max}}{Q}-1\right)Q\} ~~~~~~~~~~~~
\notag
\\
\tau^q(n) = \left\{ \begin{array}{c l}     
\{n-(q-1), ..., n-1\} & for j=1:q-1
\\
\{n, n-(Q-1), ..., n-q\} & for j=q:Q
\end{array}\right.
\end{align}
which means at each iteration only one user updates its power allocation and all the other user do not update, and this procedure sequentially continues between the users. It should be noted that due to its sequential nature, the Gauss-Seidel algorithm may require a proper scheduler in the network. Depending on the limitations of the network in terms of communication delay, one can use different update patterns, which shows the flexibility of asynchronous power control algorithm in practical situations\footnote{In the simulation part, we only use synchornous power control (i.e. Jacobi method in the sense of \cite{feyz13}) as it converges quite fast.}. In sum, as long as the transmitters are intended to update their power allocations and all the transmitters eventually become aware of other transmitters' power allocation, our power control algorithm converges under any update pattern. This result is valuable as it is applicable to other approaches that only prove the convergence of power control under Gauss-Seidel algorithm and the convergence of Jacobi algorithm is proved by conducting simulations with no theoretical analysis (see \cite{diep1}). Moreover, the convergence under asynchronous iteration mitigates the need for scheduler (which is the disadvantage of Gauss-Seidel algorithm) and accurate synchronization (which is the disadvantage of Jacobi algorithm).
\section{Simluation Results and Discussion}
In this section, we evaluate the analyses done in the previous sections and compare them to the other reference's work. We simulated a network comprised of four MIMO links that share the same band. 
 The wireless channel is comprised of an i.i.d complex Gaussian flat-fading with zero mean and unit variance for the small-scale fading
\footnote{It should be noted that the case of frequency-selective channels that need to employ MIMO-OFDM technique can be a potential subject of future work.}, 
and exponential path-loss for the large-scale fading. The additive noise is set to i.i.d complex Gaussian with zero mean and unit variance. We assume that a given scenario $N_T \times N_R$ indicates $N_T$ transmit antennas and $N_R$ receive antennas for all of the users, meaning that $N_{T_q} = N_T~\&~N_{R_q} = N_R,~\forall q \in\{1,...,Q\}$). All of the users have the same total power budget $P_q = 10 dB,~\forall q$. The distance between the $q$th transmit-receive pair is shown as $d_{qq}$ and the distance between the $r$th transmitter and the $q$th receiver $d_{rq}$. Lastly, the path-loss exponent is set to $\gamma = 2.5$.
\subsubsection{Probability of the Uniqueness of NE}
In this simulation, we want to quantify how adequately our derived conditions for the uniqueness of NE can predict the convergence of the power control game in practice. Fig. 2 shows the probability of uniqueness of NE as a function of $d_{rq}$, where $d_{qq} = 15$. Each point on the curves, is the result of the total number of times that a particular criterion is satisfied (and the game also converges to the unique NE) divided by the total number of channel realizations. 
There are 1000 total channel realizations for each point of a scenario. Each realization is done within the maximum of 100 iterations of Jacobi method. Three criteria are used for each scenario. The curves with the stars indicate whether the criterion in \cite[Condition 7]{sc1} is satisfied.
\begin{figure}
\label{fig2}
\centerline{
\includegraphics[scale = 0.57, trim= 370mm 90mm 350mm 100mm]{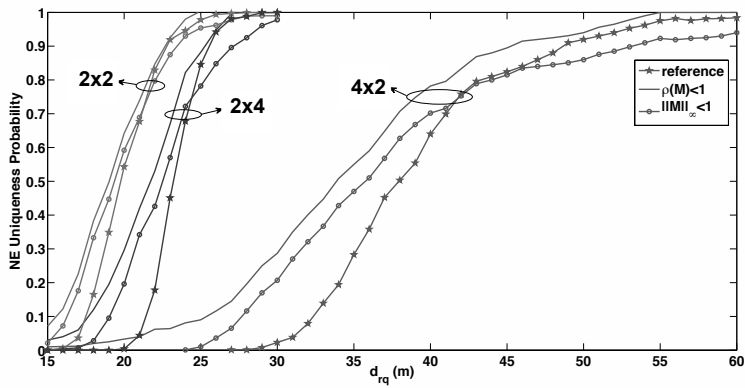}
}
\caption{\small Comparison of probability of uniqueness of NE for different number of transmit/receive antennas w.r.t. interfering distance}
\end{figure}
\begin{figure}
\label{fig3}
\centerline{
\includegraphics[scale = 0.57, trim = 370mm 90mm 350mm 100mm]{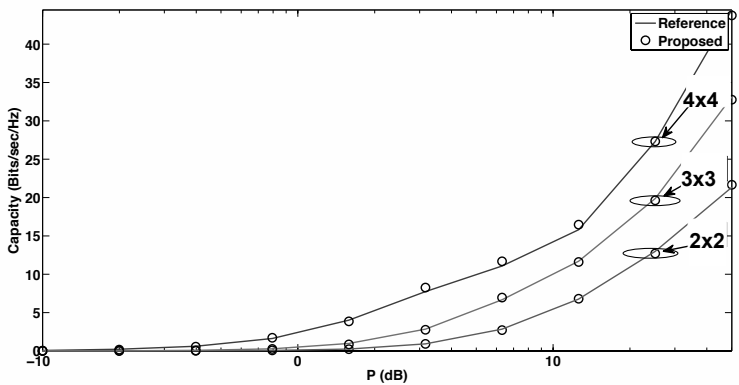}
}
\caption{\small Comparison of sum-rate for different number of transmit/receive antennas w.r.t. the total power budget of the links}
\end{figure}
 The curves with circles indicate the probability of uniqueness according to the satisfaction of \eqref{th1} or \eqref{th2}. Finally the solid curves indicate the convergence to the unique NE if $\rho(M) <1$ is satisfied. First of all, both of the convergence criteria we derived predict the uniqueness of NE more than the condition derived in \cite{sc1} for small distances because the precoding employed in our approach uses the channel information more efficiently. Furthermore, the condition $\rho(M) < 1$ predicts the uniqueness more than the condition in \eqref{th1}. As it was said, the conditions in \eqref{th1} and \eqref{th2} are derived by setting the weighting vector $v = 1$ in $||M||^v_\infty$. By setting $v = 1$, the probability of uniqueness of NE becomes less than the reference curve as the interfering distances increase\footnote{Note that the conditions derived in \cite{sc1} are not practical as well. The practical conditions derived in \cite{sc1} have worse performance loss.}. In fact, when the condition $\rho(M) < 1$ is satisfied,  the vector $v$ can have any value, but setting $v = 1$ leads to more conservative conditions that ignores some situations wherein the vector $v$ has a value other than $v = 1$. However, setting $v = 1$ leads to more practical conditions that are suitable for network designing. Lastly, for the case $N_T = 4~\&~N_R = 2$ (which is shown as $4 \times 2$ in Fig. 2), our criteria have the same trend as the criterion in \cite{sc1} does, making our proof for the case $N_T > N_R$ reasonable.
\subsubsection{Sum-Rate}
Fig. 3 shows the comparison of the sum-rate of the four MIMO links w.r.t the total power of each MIMO link. The normalized path-loss is fixed at $\left(\frac{d_{qq}}{d_{rq}}\right)^{\gamma} = 10 dB$. Each point on the curve is the mean value of 1000 independent channel realizations. For each channel realization, there are a maximum of 100 iterations of Jacobi iterative water-filling algorithm performed. Evidently, there is no difference in sum-rate between our technique and the technique used in \cite{sc1}. This is due to the fact that the Nash equilibria of the game is in general not Pareto optimal, then the selfish maximization of rate cannot be guaranteed to achieve global optimality.
\section{Conclusion}
In this paper, the concept of contractive functions is employed to derive the convergence conditions of the distributed power control in multi-user MIMO systems. The convergence conditions derived in this paper can be used in practice by the designer to create a distributed power control framework. Simulations show that the proposed criteria can predict a unique NE more than the previous work. With this advantage, the power control can converge more often, making the distributed power control a more favorable practical approach than before in terms of convergence issues. Furthermore, our proposed power control algorithm can be implemented asynchronously, which gives a noticeable flexibility to our algorithm depending on the practical limitations. Improving the achievable sum-rate of the game can be a potential future work.
\appendices
\ifCLASSOPTIONcaptionsoff
  \newpage
\fi
\bibliographystyle{IEEEtran}
\bibliography{reff2}
\end{document}